\documentstyle[epsfig]{elsart}
\begin{document}
\runauthor{Ishida}
\begin{frontmatter}
\title{$\pi^0\pi^0$ Scattering Amplitudes 
in the $\pi^-p$ Charge Exchange Process  
and $\pi^0\pi^0$ Phase Shift Analysis }

\author[MIYA]{Kunio Takamatsu}
\address[MIYA]{Miyazaki U., Gakuen-Kibanadai, Miyazaki 889-2155, Japan\\
Collaboration work with A.M.Ma,M.Y.Ishida,T.Ishida,T.Tsuru,
H.Shimizu and E135 Experimental Group at KEK\\
submitted to the proceedings of International Workshop
``$e^+e^-$ collisions from $\phi$ to $J/\psi$,'' Novosibirsk,
Russia, March 1-5, 1999.
}

\begin{abstract}
The amplitude analysis has been performed on the 
$\pi^0\pi^0$ final state obtained in the $\pi^-p$ charge exchange 
process.
The $\pi^0\pi^0$ scattering amplitudes have been obtained for the S 
and D waves by the Chew-Low extrapolation
and the partial wave analysis. 
Breit Wigner parameters have been obtained for 
$f_0(1370)$ and $f_2(1270)$. 
I=0 S wave $\pi^0\pi^0$ 
scattering phase shift has been obtained
below $K\bar K$ threshold. They agree well 
with the $\pi^+\pi^-$ standard phase shift 
data below 650GeV and deviate by 
about 10 degrees from the standard data above 650MeV. 
They show a different behavior  from those of Cason and 
others. 
The $\pi^0\pi^0$  phase shift data have been 
analyzed by the IA method. 
Resonance parameters have been obtained
to be 
$m_\sigma =588 \pm 12$MeV and $\Gamma_\sigma =281\pm 25$MeV with 
$r_c=2.76\pm 0.15$GeV$^{-1}$. They are excellently in agreement 
with those obtained in the reanlysis on the $\pi^+\pi^-$phase 
shift data.
\end{abstract}
\end{frontmatter}

%
{\bf Introduction}:\ \ \ \ 
The scalar meson $\sigma$, a chiral partner of pion as 
the Nambu-Goldstone Boson has long been 
expected\cite{ref1} to be found. A mass of $\sigma$ is predicted 
to be twice of that of the constituent quark mass, $m_q$ in 
the NJL model, the low energy effective theory of QCD, 
i.e. $2m_q=500-700$ MeV. It was so pity that the existence of it 
had long been rejected in the analysis\cite{ref3} of the $\pi\pi$ 
scattering phase shift data. The situation has been 
improved for its existence in recent studies\cite{ref4,ref5} 
of re-analyses of the $\pi\pi$ phase shift data.\cite{ref8} 
$\sigma$ has revived 
on the PDG table\cite{ref7} after 20 years of vanishing. 
S.Ishida et al.\cite{ref5} have, especially, shown the clear 
evidence of $\sigma$ in their reanalysis of the $\pi\pi$ phase 
shift data by the interference amplitude (IA) method with 
introduction of a negative background phase in the analysis. 
The introduction
 of the negative background is not a simply 
symptomatic treatment for the analysis, but has reasoning on the 
experimental fact that the S wave I=2 $\pi\pi$ scattering phase 
shift\cite{ref8} is negative and has also reasoning 
in the theoretical consideration\cite{ref5,ref10} that the negative
background phase has an origin in the 
compensating $\lambda\phi^4$ contact 
interaction term requested by the current algebra and PCAC. \\
\hspace*{1cm}It might be vital important to see $\sigma$ in the 
production process, as well. $\sigma$ has been shown\cite{ref11} in 
the analysis of the $\pi^0\pi^0$ system produced in the $pp$ central 
collision 
process at 450 GeV/c performed by the NA12/2 experiment at CERN. 
The variant mass and width method\cite{refVMW} were used for the analysis.
The similar analysis\cite{ref13} has been perfomed 
on the $\pi\pi$ final state produced in the $J/\Psi$ decay 
into $\omega\pi^+\pi^-$ obtained by the DM2 
collaboration\cite{ref12} at DCI. The $\pi\pi$ mass distribution as
 well as angular distributions have been well reproduced in the 
analysis. \\
\hspace*{1cm}A sizable accumulation of $\pi\pi$ events produced in 
the central collision process has been treated\cite{ref14}, so far,
 to be of a background process according to the conclusion from the
 traditional analyses\cite{ref3} of $\pi\pi$ phase shift data.
Though the comments\cite{ref15} were stated on the preliminary 
results\cite{ref16} of the analysis of the $\pi^0\pi^0$ scalar 
state below 1GeV in the concluding remarks at Hadron'95, they lose
 now their physics ground. The comment\cite{ref17} should also be 
retracted, which was given in the summary talk at Hadron'97 to the 
results of the reanalysis\cite{ref5} on the $\pi^+\pi^-$ phase
 shift data, without notice of the essential points of the analysis.
Discussions more
 in detail can be found in the reports at Hadron'97 and
 at this workshop\cite{ref18}.\\
\hspace*{1cm}It can be seen apparent deviations of the $\pi^0\pi^0$ 
phase shift data\cite{ref20} from the $\pi\pi$ phase 
shift data\cite{ref8}(the standard $\pi\pi$ phase shift data, so called).
 Meanwhile, we performed an 
amplitude analysis on the $\pi^0\pi^0$ final state produced 
in the $\pi^-p$ charge exchange reaction, 
$\pi^-p \rightarrow\pi^0\pi^0n$ at KEK. $\pi^0\pi^0$ scattering 
phase shifts 
have been obtained and analyzed. We present here the results of 
our amplitude analysis on the $\pi^0\pi^0$ final state and of 
analysis of the scattering phase shifts with the IA method.

{\bf $\pi^0\pi^0$ scattering amplitudes in the $\pi^-p$ charge 
exchange process}:\ \ \ \ 
An amplitude analysis has been performed on the $\pi^0\pi^0$  final
 state produced in the $\pi^-p$ charge exchange process, 
$\pi^-p \rightarrow\pi^0\pi^0n$ at 8.95 GeV/c. The data were 
collected by the Benkei spectrometer of the E135 experiment at 
12 GeV proton synchrotron at KEK (Tsukuba). The Benkei spectrometer
 had a high resolution performance with wide geometrical acceptance
 for charged particles and gammas. The neutral events of four gammas 
with no charged particle have been used for the present analysis. 
The gammas produced in a liquid hydrogen target hit by the negative
 pion beam were detected by a total absorption hodoscope 
spectrometer consisted with active converters (AC) and main 
radiators (MR) of lead glass. Positions of gammas were determined 
by tracks recorded in wire chamber planes placed between AC and 
MR. Recoil neutrons were not detected. Reader can refer details of 
the system elsewhere\cite{ref21}.
\begin{figure}[t]
 \epsfysize=4.0 cm
 \centerline{\epsffile{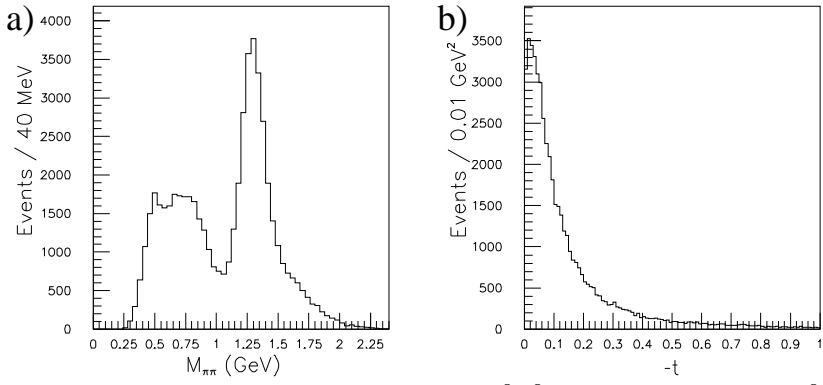}}
 \caption{(a) Effective mass distribution of the $\pi^0\pi^0$ system 
in $\pi^-p \rightarrow\pi^0\pi^0n$. Data are corrected by the  
acceptance.
(b) t distribution of $\pi^-p \rightarrow\pi^0\pi^0n$ .
}
\label{fig2}
\end{figure}
%
%
%
Reconstructed $\pi^0$ signals were selected in the region,
$|M_{2\gamma}-M_{\pi^0\pi^0}|\leq 40$MeV. The mass resolution for
$\pi^0$ detection was 13MeV. The missing mass squared against 
$\pi^0\pi^0$ was selected in the region between 0.3 and 1.21
GeV$^2$. The effective mass distribution of the $\pi^0\pi^0$ system
 is shown in Fig.1a). A broad peak is seen below 1GeV with a clear 
peak around 1.3GeV. The t distribution in Fig.1b) shows that the one 
pion exchange dominates in the process. We may write the cross 
section for the one pion exchange process, as follows,\\
\begin{eqnarray}
\frac{d\sigma}{dm_{\pi\pi}d{\rm cos}\ \theta dt}
 & \sim &  \frac{m_{\pi\pi}^2}{|p_1|}\frac{4m_N^2-t}{(m_\pi^2-t)^2}
T_{\pi\pi}^2(m_{\pi\pi}^2,{\rm cos}\ \theta ,t)
\label{eq1}
\end{eqnarray}
where $|p_1| = \sqrt{m_{\pi\pi}^2/4-m_\pi^2}$ is the pion momentum in 
the $\pi\pi$ center of mass system. $\theta$ is the scattering 
angle of the $\pi\pi$ system in the 
$\pi^+\pi^-\rightarrow\pi^0\pi^0$ scattering or the azimuth angle 
in the G-J frame. $T_{\pi\pi}$ is the off mass shell 
scattering amplitude of the pion.\\
\hspace*{1cm}The $\pi^0\pi^0$ system is in a state with even 
angular momentum and even isospin. The on mass shell scattering 
amplitude is written with S and D waves, as follows,\\
\begin{eqnarray}
T_{\pi\pi}(m_{\pi\pi}^2,{\rm cos}\ \theta ,m_\pi^2)
 &=&  A_S+A_D\times\sqrt{5}\times\frac{3{\rm cos}^2\ \theta -1}{2} 
\label{eq2}
\end{eqnarray}
where $A_S$ and $A_D$ are the on mass shell amplitudes for S and D 
waves, respectively. A linear form is used for the 
extrapolation\cite{ref22} of the off mass shell amplitude to the 
on mass shell one.\\
\begin{eqnarray}
T_{\pi\pi}^2(m_{\pi\pi}^2,{\rm cos}\ \theta ,t)
 &=&  (1+\alpha (m_\pi^2 -t))
T_{\pi\pi}^2(m_{\pi\pi}^2,{\rm cos}\ \theta ,m_\pi^2)
\label{eq3}
\end{eqnarray}
where $\alpha$ is the extrapolation coefficient. 
Then, the cross section becomes\\
\begin{eqnarray}
\sigma \sim 
\frac{m_{\pi\pi}^2}{|p_1|}\ \frac{4m_N^2-t}{(m_\pi^2-t)^2} 
 & &   (1+\alpha (m_\pi^2 -t))   \\
\times \left(  A_S^2+ 5 A_D^2    
      \left(\frac{3{\rm cos}^2\ \theta -1}{2}\right)^2\right.
 & +&  \left. 2\sqrt{5}\   |A_S|\  |A_D|\ 
\frac{3{\rm cos}^2\ \theta -1}{2}\  
{\rm cos}\ \delta\right) 
\nonumber
\label{eq4}
\end{eqnarray}
where $\delta$ is the relative phase between $A_S$ and $A_D$. 
\begin{figure}[t]
 \epsfysize=10.0 cm
 \centerline{\epsffile{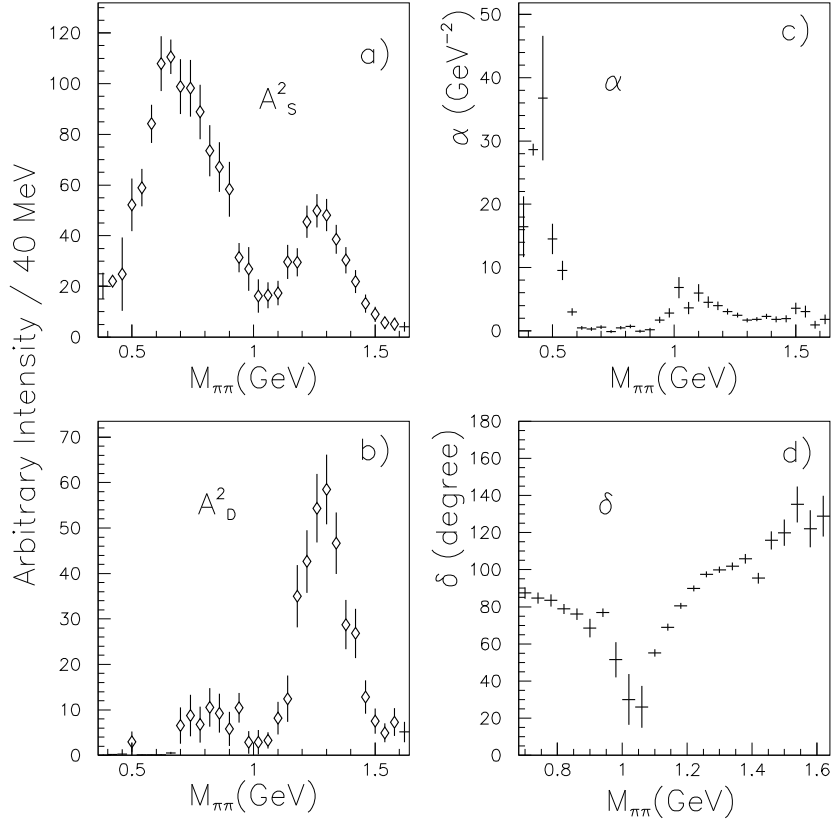}}
 \caption{Results of the partial wave analysis. 
(a) Intensity distribution of the S wave amplitude square. 
(b) Intensity distribution of the D wave amplitude square. 
(c) Extrapolation coefficient $\alpha$.
(d) Relative phase $\delta$ between S and D wave amplitudes,
$\delta =\phi_D-\phi_S$ below 1.0GeV and $\delta =\phi_S-\phi_D$ 
above 1GeV.}
\label{fig5}
\end{figure}
The results of PWA are shown in Fig.2a)-d). A broad peak is seen 
below 1GeV in the S wave intensity distribution ( Fig.2a)). 
A rapid fall around 0.9GeV may be due to the interference of the 
S wave with $f_0(980)$ which appears as a sharp variation of the 
phase motion in Fig.2d). A clear peak around 1.3GeV in Fig.2a) 
may correspond to $f_0(1370)$. No structure is seen 
around 1.5GeV. A clear peak of $f_2(1270)$ is seen in the D wave 
intensity distribution (Fig.2b)). A week but significant intensities
are seen below 1GeV. It might be interesting to note that the 
similar events are recognized in the intensity distributions of 
the $\pi^+\pi^-$ and $\pi^0\pi^0$ final states produced in the $pp$
 central collision production\cite{ref23}.
%
%
The Breit-Wigner amplitude parameters are obtained for the peaks 
around 1.3GeV in the S and D waves with the relativistic form. 
Parameters obtained for masses and widths of $f_0(1370)$ and $f_2(1270)$ 
are as follows,
\begin{eqnarray}
M_{f_0(1370)} &=& 1278\pm 5\ \ {\rm MeV},
  \ \ \ \ \Gamma_{f_0(1370)} = 197\pm 8\ \ {\rm MeV},\nonumber \\
M_{f_2(1270)} &=& 1286\pm 7\ \ {\rm MeV},
  \ \ \ \ \Gamma_{f_2(1270)} = 161\pm 14\ \ {\rm MeV}.
\end{eqnarray}
\begin{figure}[t]
 \epsfysize=7.0 cm
 \centerline{\epsffile{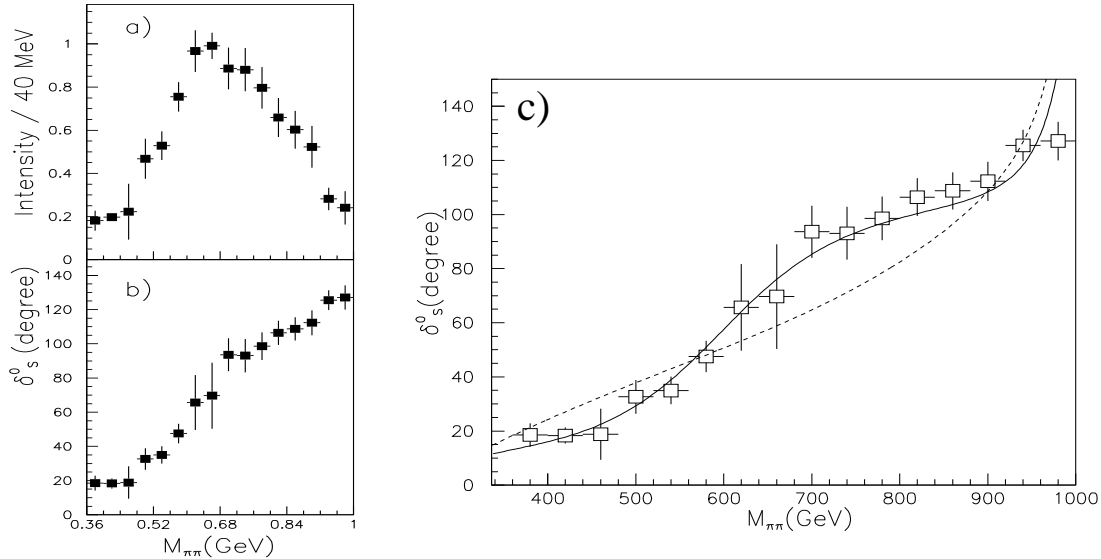}}
 \caption{ (a) Normalized intensity distribution of S wave 
amplitude square below 1.0 GeV. 
(b) I=0, S wave $\pi^+\pi^-\rightarrow\pi^0\pi^0$ scattering phase 
shift $\delta_S^0$ below $K\bar K$ threshold. 
(c) The same for  (b). Solid line is the 
result of fitting with the IA method introducing the negative phase
 background $\delta_{BG}$.
Dotted line shows the result with $r_c=0$.
}
\label{fig7}
\end{figure}
The $\pi^0\pi^0$ scattering amplitudes below $K\bar K$ 
threshold are expressed by the S wave I=0 and I=2 phase shifts,
$\delta_S^0$ and $\delta_S^2$, respectively, as follows,
$A_S^2\sim {\rm sin}^2(\delta_S^0-\delta_S^2)$.
The $\delta_S^0-\delta_S^2$ distribution is deduced 
by normalizing the maximum value of $|A_S|^2$ to be 1. 
The experimental $\delta_S^2$ is shown to be well reproduced 
by the hard core type phase shift,  $\delta_S^2=-r_{c2}|p_1|$, with core
radius $r_{c2}=0.17{\rm fm}=0.87{\rm GeV}^{-1}$.\cite{ref5}
By using this $\delta_S^2$ we can obtain
$\delta_S^0$, shown in Fig.3. $\delta_S^0$ values
 are in good agreement with the $\pi^+\pi^-$ standard phase shift 
data below 650MeV. They appear somewhat higher by about 10 degrees 
than the standard phase shift
above 650MeV . They are different as a whole from those of Cason 
and others\cite{ref20}, which have currently been used for phase 
shift analyses.
Recently, 
Kaminski and others\cite{ref24} have obtained $\pi^+\pi^-$ phase 
shifts in their reanalysis of the polization data. Their up-flat 
solution shows apparent deviation from ours.  
The phase difference $\delta_S^0-\delta_S^2$ at the 
neutral Kaon mass is obtained in our analysis with the value 
42.5$\pm$3$^\circ$, which is consistent with the prediction from the CP 
violation parameters in the K-decay,\cite{ref28} 40.6$\pm$3$^\circ$.

%
{\bf $\pi^0\pi^0$ phase shift analysis}:\ \ \ \ 
$\pi^0\pi^0$ phase shifts data are analyzed by the IA 
method\cite{ref5}, which describes the process with a 
few physically clear parameters.
S matrix is written with phase shift, $\delta (s)$ and 
scattering amplitude, $a(s)$, as follows,
$S=e^{2i\delta (s)}=1+2ia(s)$.
$\delta (s)$ is the sum of phase shifts come from resonances 
concerning, $\delta_R$'s and background phase shift, 
$\delta_{BG}$ below $K\bar K$ threshold. $f_0(980)$ and
$\sigma$ are taken as resonances. 
Then, $\delta (s) = \delta_R + \delta_{BG}
=\delta_{f_0} + \delta_\sigma +\delta_{BG}$. The relativistic 
Breit-Wigner is taken for $a(s)$, 
$a(s)=\sqrt{s}\Gamma_R(s)/(m_R^2-s-i\sqrt{s}\Gamma_R(s))$. 
Total S matrix is 
written as, $S=S_RS_{BG}=S_{f_0}S_\sigma S_{BG}$. 
The unitarity condition for $S$ is satisfied by each $S$ matrix,  
automatically. The negative phases are taken for $\delta_{BG}$ which are 
expressed by the hard core, $\delta_{B}=-r_c|p_1|$. The solid line 
presented in Fig.3c) is the best fitted curve obtained below 
$K\bar K$ threshold by fitting with parameters, resonant mass 
$m_\sigma$, resonant width $\Gamma_\sigma$ and hard core radius 
$r_c$. The data are excellently reproduced by the curve. The $\chi^2$
is obtained with $\chi^2/n_{n.o.f}=20.4/12$.
Parameters obtained are,
\begin{eqnarray}
m_\sigma &=& 588\pm 12{\rm MeV},\ \Gamma_\sigma =281\pm 25{\rm MeV}
\  {\rm and}\  r_c=2.76\pm 0.15{\rm GeV}^{-1}
.
\end{eqnarray}
These values are in good agreement with those which we have obtained 
in the reanalysis on the standard $\pi^+\pi^-$
phase shift data. 
A dotted line in FiG.3c) shows also the result of
 fitting with $r_c=0$ (i.e. without negative background), that is 
the same with the case of the traditional analysis. 
The $\chi^2/n_{n.o.f}$ is obtained with $85.0/13$,
worse than our best fit.
The parameter obtained with $r_c=0$
are $m_{``\sigma ''} =890\pm 16{\rm MeV}\ \  {\rm and}\ \ 
\Gamma_{``\sigma ''} =618\pm 51{\rm MeV}.$
\end{document}